\documentclass[superscriptaddress,onecolumn,aps,amsmath,amssymb,prb,preprint]{revtex4}
\usepackage{graphicx}
\usepackage{dcolumn}
\usepackage{bm}
\usepackage{times}
\usepackage{amsmath}
\usepackage{epsfig}
\usepackage{feynmp}
\usepackage{epstopdf}
\usepackage{setspace}

\begin{document}
\setlength{\unitlength}{1mm}
\title{Large-$N$ estimates of universal amplitudes of the $\mathbb{CP}^{N-1}$ theory and comparison with the JQ model}
\author{Ribhu K. Kaul}
\affiliation{Department of Physics, Harvard University, Cambridge, MA 02138.}
\author{Roger G. Melko}
\affiliation{Department of Physics and Astronomy, University of Waterloo, Ontario, N2L3G1, Canada}
\date{\today}
\begin{abstract}
We present computations of certain finite-size scaling functions and universal amplitude ratios in the large-$N$ limit of the $\mathbb{CP}^{N-1}$ field theory. We pay particular attention to the uniform susceptibility, the spin stiffness and the specific heat. 
Field theoretic arguments have shown that the long-wavelength description of the phase transition between the N\'eel and valence bond solid states in square lattice $S=1/2$ anti-ferromagnets is expected to be the non-compact $\mathbb{CP}^1$ field theory. 
We provide a detailed comparison between our field theoretic calculations and quantum Monte Carlo data close to the N\'eel-VBS transition on a $S=1/2$ square-lattice model with competing four-spin interactions (the JQ model). 
\end{abstract}
\maketitle

\section{Introduction}

The emergence of relativistic gauge theoretic descriptions of complex condensed matter systems at long wavelengths is an exciting theme of modern physics. It has now been established, as a proof of principle, that such descriptions can arise in suitably engineered lattice models of bosons~\cite{motsen}. The search for experimentally motivated microscopic models where such emergent phenomena may be observed is clearly of great interest, as this will pave the way to the detection and realization of these fascinating phenomena in experiment (e.g. in ultra-cold atoms in optical potential), and may hold the key to understanding the myriad mysterious properties of a growing number of strongly correlated materials.  The most famous class of such materials are the cuprates, which begin as insulating $S=1/2$ square lattice N\'eel ordered anti-ferromagnets, and then evolve into high-temperature superconductors on the introduction of a small density of charge carriers. It has long been held that a complete understanding of the insulating quantum anti-ferromagnet and particularly its transition to a paramagnetic phase from the N\'eel state holds the key to the cuprate mystery~\cite{Anderson,LNW}. 

Motivated primarily by this notion, a large body of work on the paramagnetic phases of frustrated, square-lattice, $S=1/2$, $SU(2)$ symmetric anti-ferromagnetic spin models has developed. Field theoretic work~\cite{rs} established that the paramagnet that arises by condensing topological defects of the N\'eel order parameter is a  valence-bond solid (VBS). Consistent with this field theoretic argument, exact diagonalization of a $S=1/2$ square lattice anti-ferromagnetic spin model with a ring exchange on clusters with upto 40 spins have also found VBS phases on the destruction of N\'eel order~\cite{laeuchli} as do series expansion studies on the $J_1-J_2$~\cite{singh}. There appears to be still some uncertainties on the $J_1-J_2$ model which has been reviewed~\cite{lhuillier}.  Recently, however an exciting step in this direction has been achieved: it has been possible to study the destruction of N\'eel order and appearance of the VBS phase on lattices with up to $10~000$ spins~\cite{melkokaul} in an unbiased way in the so-called JQ model~\cite{JQ_1} introduced by Sandvik by using sign-problem free quantum Monte Carlo (QMC) techniques: 
\begin{equation}
\label{eq:jq}
H_{\rm JQ}=J\sum_{\langle ij \rangle} {\bf S}_i \cdot {\bf S}_j
-Q\sum_{\langle ijkl \rangle} ({\bf S}_i \cdot {\bf S}_j-\frac{1}{4})
({\bf S}_k \cdot {\bf S}_l-\frac{1}{4}),
\end{equation}
where the first term is summed on nearest neighbor bonds of the square lattice and the second term is summed over plaquettes allowing dot products only on nearest neighbor bonds.  The JQ model harbors a N\'eel phase at $J>0,Q=0$ and a VBS phase at $J=0,Q>0$ and a transition between them at $J/Q\approx0.040, Q>0$.  In Ref.~\cite{JQ_1,melkokaul}, the analysis of the QMC data provided evidence for a continuous transition. Subsequent QMC work has claimed a very weak first order transition using a ``flowgram'' analysis, although no detectable discontinuity in any physical quantity has been observed~\cite{shailesh}. Although the nature of this transition is currently under debate, further numerical work will likely be able to sort this out unambiguously, thanks to the absence of a sign problem in this model. Regardless of the fate of the transition at arbitrarily long length scales, there is clear evidence~\cite{JQ_1,melkokaul,shailesh} for very large correlation lengths and the associated scaling behavior on the relatively large intermediate length scales that have been simulated. In this paper, we take the natural point of view that this (possibly approximate) scaling behavior is the result of a nearby fixed point. An interesting challenge that then immediately arises is to identify the fixed point that gives rise to the observed scaling. 

A candidate for the fixed point was predicted in an extension~\cite{DQCP12} of the field theoretic argument alluded to earlier. It has been shown that the long-wavelength description of the transition from the N\'eel state to the VBS state should be written in terms of the $\mathbb{CP}^1$ field theory of two complex bosonic spinons $z_\alpha$ interacting with a gauge field, $A_\mu$: 
\begin{equation}
\label{eq:cp1}
S_{\mathbb{CP}^{N-1}} = \sum_{\alpha=1}^N\int d^2r d\tau  |(\partial_\mu - i A_\mu)z_\alpha|^2
\end{equation}
with the constraint $\sum_\alpha |z_\alpha|^2=1$, at $N=2$. An analysis of the Berry phases of the topological defects leads to the conclusion that only quadrupled monopoles of $A_\mu$ are permitted in the continuum limit~\cite{haldane88,rs}. At $N=1$ duality transformations~\cite{dasgupta} establish that the quadrupled monopoled are irrelevant at the continuous transition of this field theory; they are hence also almost certainly irrelevant at the putative critical point at $N=2$. This leads to the remarkable conclusion that the long wavelength description of the N\'eel-VBS transition is described by Eq.~(\ref{eq:cp1}) at $N=2$ with a {\em non-compact} gauge field. It has been argued by Motrunich and Vishwanath~\cite{motrunich} that the critical point of the non-compact field theory at $N=2$ belongs to a new `deconfined' universality class, distinct from the $O(3)$ universality class obtained when the gauge field is compact. Another study has also found evidence for the new universality class~\cite{flavio}. An accurate numerical estimate of the critical exponents and universal amplitudes of this new universality class are currently lacking, since Monte Carlo studies have been restricted to relatively small lattices. Numerical studies of both the non-compact~\cite{motrunich,kamalmurthy} and compact~\cite{ichinose} $\mathbb{CP}^{N-1}$ model are available. An alternate approach to a quantitative study, which we follow here, is to construct an expansion around the $N=\infty$ fixed point. Indeed the $N=\infty$ fixed point is stable at large but finite $N$, and universal quantities may be computed in a $1/N$ expansion and extrapolated to $N=2$. Our study {\em assumes} that there is a fixed point at $N=2$ that is continuously connected to the large-$N$ fixed point, as is true for instance in the $O(N)$ model; we however cannot prove that this is true. We note that our focus here is exclusively on the full $SU(2)$ symmetric case; there has been extensive work~\cite{u1many} on a $U(1)$ deformation of Eq.~(\ref{eq:cp1}), which was predicted~\cite{DQCP12} to be the critical theory of a quantum transition between a superfluid and a valence bond solid in $U(1)$ symmetric spin models~\cite{sandvikJK}, however our results do not apply to this case. 

The universality class of a fixed point is characterized by the values of the critical exponents, amplitudes and scaling functions close to the transition. 
We focus here on certain quantities that are associated with susceptibilities of conserved charges. We provide novel computations of (a) finite-size scaling functions for these quantities and (b) ratios of the universal amplitudes in the large-N expansion. 
We find reasonable agreement with QMC data on the JQ model, thus providing some support for the hypothesis that the scaling behavior observed close to the N\'eel-VBS transition in the JQ model is due to a proximity to the fixed point of the $\mathbb{CP}^1$ model. We note however that our calculations are only a first step and a fully convincing demonstration would require a comparison with a numerical study of an appropriate lattice discretization of the non-compact $\mathbb{CP}^1$ field theory, i.e. working directly at $N=2$.

\section{Large-N Formalism}
\label{sec:lgN}
We are interested in studying two-dimensional quantum anti-ferromagnets, that are described by Eq.~(\ref{eq:cp1}), at finite-temperatures. This clearly requires a study of the field theory in a slab geometry where the extent in the third direction is proportional to $1/T$. We will be interested also in the effect of finite spatial extent of linear dimension, $L$. We describe the formalism used for these calculations in this section.

We begin with the resolution of the constraint on $z_\alpha$, by introducing a real field $\lambda$, which acts as a Lagrange multiplier at each point of space and time. 
\begin{equation}
S_b =\frac{1}{g} \int d^2 r d\tau \left [ |(\partial_\mu - iA_\mu) z_\alpha|^2 - i\lambda (|z_\alpha|^2-N)\right ]
\end{equation}  
Note that the integration on $\tau$ is carried out from $0$ to $1/T$ and the corresponding Fourier transform consists of the Matsubara modes $\omega_{n_\tau}=2\pi n_{\tau}T$, and likewise the spatial integral is from $0$ to $L$ with corresponding Matsubara modes labeled by $n_x$ and $n_y$ . In the limit $N=\infty$ the gauge field drops out and $\lambda$ takes on a uniform saddle point value. 
We can compute all quantities of interest from the free energy at criticality. We will organize its large-N expansion as,
\begin{equation}
\mathcal{F} = N f^{0} + f^{1\lambda} + f^{1A}
\end{equation}

At $N=\infty$, the problem reduces to $N$ free complex scalar fields,
\begin{equation}
\label{eq:freebox}
f^{0}=\frac{T}{2L^2} \sum_{n_\tau,n_x,n_y, \theta=\pm 1}\ln\left[ \left( \omega_{n_\tau}-\frac{\theta H}{2}\right)^2 + k^2_{n_x}+k^2_{n_y}+m^2_{\rm box}\right ] - \frac{m_{\rm box}^2}{g_c}
\end{equation}
where we have included a convenient magnetic field $H$, which enables a computation of the uniform susceptibility, $\chi_u=\frac{\partial^2 \mathcal{F}}{\partial H^2}$.

We now turn to an overview of the computation of the free energy at next order, i.e. $f^{1\lambda}$ and $f^{1A}$. We can organize the effective action for the $\lambda$ and $A$ fluctuations as.
\begin{eqnarray}
\label{eq:SeffTfin}
\mathcal{S}_{A,\lambda} = \frac{T}{2L^2} \sum_{\epsilon_n,k_x,k_y}
\left[ \left( k_i A_\tau - \epsilon_n A_i \right)^2 \frac{D_{1} ({\bf k},
\epsilon_n)}{{\bf k}^2} + A_i A_j \left( \delta_{ij} - \frac{k_i k_j}{{\bf k}^2}
\right) D_{2} ({\bf k}, \epsilon_n) + \Pi_\lambda \lambda \lambda\right] .
\end{eqnarray}
We will avoid explicit details of the computations of $D_{1,2}$ and $\Pi_\lambda$ here, since they have already been presented Ref.~\onlinecite{kaulsachdev}. In terms of the functions $D_{1,2}$ and $\Pi_\lambda$ the free energy is,
\begin{eqnarray}
\label{eq:free_en1}
f^{1\lambda} &=& \frac{T}{2L^2}\sum_{\omega_n,k_x,k_y}  \ln\left (\Pi_\lambda  \right ) \\
f^{1A} &=& \frac{T}{2L^2}\sum_{\omega_n,k_x,k_y}  \ln\left (D_1\left[ D_2+\frac{\epsilon_n^2}{k^2}D_1 \right] \right ) 
\end{eqnarray}
These expressions are useful to compute the $1/N$ corrections to the Wilson Ratio in Sec.~\ref{sec:wilson}

The focus of this paper is on a computation of universal amplitudes associated with susceptibilities of conserved quantities. Before turning to these calculations, for completeness, we briefly discuss two critical exponents at the transition: $\nu$, the correlation length exponent and $\eta$  the anomalous dimension of the N\'eel field, $\vec{n}=z^* \vec{\sigma}z$ at criticality.
Large-$N$ computations for these quantities produce the results~\cite{hlm,ikk,kaulsachdev},
\begin{equation}
\eta=1-\frac{32}{\pi^2 N}; ~~~~\nu = 1 - \frac{48}{\pi^2N}
\end{equation}
Note that these values become negative and unphysical for the case of interest ($N=2$), so they do not provide any quantitative information for $N=2$. Presumably, the higher order corrections are large and cannot be neglected. It is interesting however to note that the anomalous dimension, $\eta$ is expected to be large, since at $N=\infty$ it approaches $1$, which is a result of the fact the N\'eel order parameter is not the field that renders the action quadratic at mean-field level. This observation agrees qualitatively with the QMC data on the JQ model,~\cite{melkokaul} where a scaling analysis close to the phase transition found an anomalous dimension, $\eta\approx0.35$, that was almost an order of magnitude larger than that of the conventional $O(3)$ universality class. Reassuringly the first term of order $1/N$ is of the correct sign (negative) correcting the the $N=\infty$ result, $\eta=1$, in the correct direction.  We now turn to large-N computations of certain scaling functions and amplitudes ratios.

\section{Finite Size Scaling functions for $\mathbb{CP}^{N-1}$ model at $N=\infty$} 

In this Section we will work only at $N=\infty$, but will study arbitrary values of the parameter $LT$.
We need to first extremize $f^0$ to obtain the large-$N$ mass equation in a box:
\begin{equation}
\label{eq:masseq}
\frac{T}{L_x L_y}\sum_{n_x, n_y, n_\tau}\frac{1}{\omega^2_{n_\tau}+ k^2_{n_x}+ k^2_{n_y}+m^2_{\rm box}}=\int\frac{d^3p}{8\pi^3}\frac{1}{p^2}
\end{equation}
where $\omega_n= 2\pi T n$ and $k_n=\frac{2\pi}{L} n$ and this equation has to be solved self-consistently to obtain the saddle point value, $m_{\rm box}$.

\begin{figure}[]
\includegraphics[width=3.2in]{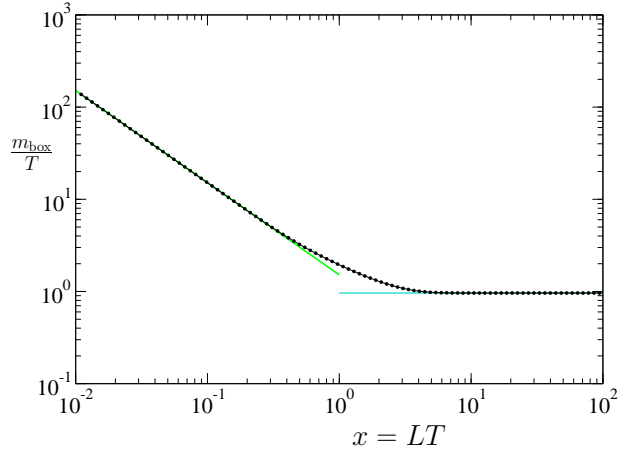}
\caption{ \label{fig:lgNmass} Large-$N$ saddle-point value of the mass as a function of the parameter, $x=LT$. The asymptotes for $x\ll 1$ and $x\gg 1$ are plotted for comparison.}
\end{figure}

We can simplify Eq.~(\ref{eq:masseq}) by using the Poisson summation formula,
\begin{equation}
\frac{T}{L_x L_y}\sum_{n_x, n_y, n_\tau}\frac{1}{\omega^2_{n_\tau}+ k^2_{n_x}+ k^2_{n_y}+m_{\rm box}} = \sum_{{\bf N}\in \mathbb{Z}^3}\int \frac{d^3p}{8\pi^3}\frac{e^{i p_i N_i L_i}}{p^2 + m_{\rm box}^2}
\end{equation}
where $i=x,y,\tau$ and $L_\tau=\beta$. Taking the ${\bf N}=0$ term on the RHS of Eq.~(\ref{eq:masseq}), we obtain
\begin{equation}
\sum_{{\bf N}\neq 0}\int \frac{d^3p}{8\pi^3}\frac{e^{i p_i N_i L_i}}{p^2 + m_{\rm box}^2}=\frac{m_{\rm box}}{4\pi}
\end{equation}
The integral on the LHS can be evaluated exactly in 3 dimensions, giving us the following simple non-linear equation for $m=\frac{ m_{\rm box}}{T}$:
\begin{equation}
\sum_{{\bf N}\neq 0}\frac{e^{-m r_{\bf N}}}{r_{\bf N}} = m,
\end{equation} 
where $r_{\bf N}= \sqrt{(N_x^2 + N_y^2)x^2 + N_\tau^2}$, where $x=L T$. A solution of this non-linear equation is shown in Fig.~\ref{fig:lgNmass} where the self-consistent mass is plotted as a function of the scaling parameter $x=LT$. When $x \gg 1$, we know $m(x)= 2 \log (\frac{\sqrt{5}+1}{2})\approx0.962424$ and for $x \ll 1$, we know $m(x)\approx 1.51196/\zeta$, these asymptotes have also been plotted for comparison.

We now compute the spin stiffness, $\rho_s$, the uniform susceptibility, $\chi_u$, and the specific heat, $C_V$, for the $\mathbb{CP}^{ N-1}$ model, when it is placed in a box of linear dimension $L$ and at temperature $T$. We note that due to the absence of any anomalous scaling dimension in $\rho_s$ and $\chi_u$, the scaling forms are completely universal. In this section we restrict ourselves to the $N=\infty$ case.

 In the scaling limit, proximate to the critical point, we can quite generally write,
\begin{eqnarray}
\label{eq:scaling}
L^2T\chi_u &=& \mathbb{Z}(\frac{L^z T}{c},t L^{1/\nu})\\
\frac{\rho_s}{T} &=& \mathbb{Y}(\frac{L^z T}{c},t L^{1/\nu})\\
\frac{L^2C_V}{T}&=& \mathbb{X}(\frac{L^z T}{c},t L^{1/\nu})
\end{eqnarray}
where $t$ measure deviations from the critical coupling, $z$ is the dynamic exponent and $c$ is a non-universal velocity. In our large$-N$ calculations, we set $c=1$ and hence ignore its presence; in our QMC calculations on the other hand, $c$ is determined by the details of our JQ model and we have to estimate it from numerical simulations.

Now that we have calculated the value of the mass parameter, we can calculate the $N=\infty$ value of $\chi_u$ and $\rho_s$ as a function of $x=LT$.
\begin{eqnarray}
\chi^\infty_u&=&\frac{NT}{2L^2} \sum_{n_\tau,n_x,n_y} \left [ \frac{1}{k_{n_x}^2+k_{n_y}^2+\omega_{n_\tau}^2+m_{\rm box}^2} - \frac{2\omega^2_{n_\tau}}{(k_{n_x}^2+k_{n_y}^2+\omega_{n_\tau}^2+m_{\rm box}^2)^2} \right ]
\end{eqnarray}
Completing the Matsubara sum and re-writing in units of $T$, we find,
\begin{equation}
\label{eq:chiuinfty}
\frac{\chi^\infty_u}{T}=\frac{N}{8x^2}\sum_{n_x,n_y}\frac{1}{\sinh\left[ \frac{\sqrt{(\frac{2\pi}{x})^2(n_x^2+n_y^2)+m^2 }}{2}\right] }
\end{equation}
The sums on $n_x$ and $n_y$ converge fairly rapidly and we can evaluate them by simply introducing cutoffs in the sums. We know from simple hyper-scaling laws that $\mathbb{Z}(x\rightarrow \infty, 0)=\mathcal{A}_\chi x^2$. In order to evaluate $\mathcal{A}_\chi$ we convert the sums in Eq.~(\ref{eq:chiuinfty}) into integrals and the RHS becomes $\mathcal{A}^0_\chi/N=0.0856271$, the mean-field value of the universal amplitude.

\begin{figure}[]
\includegraphics[width=6.in]{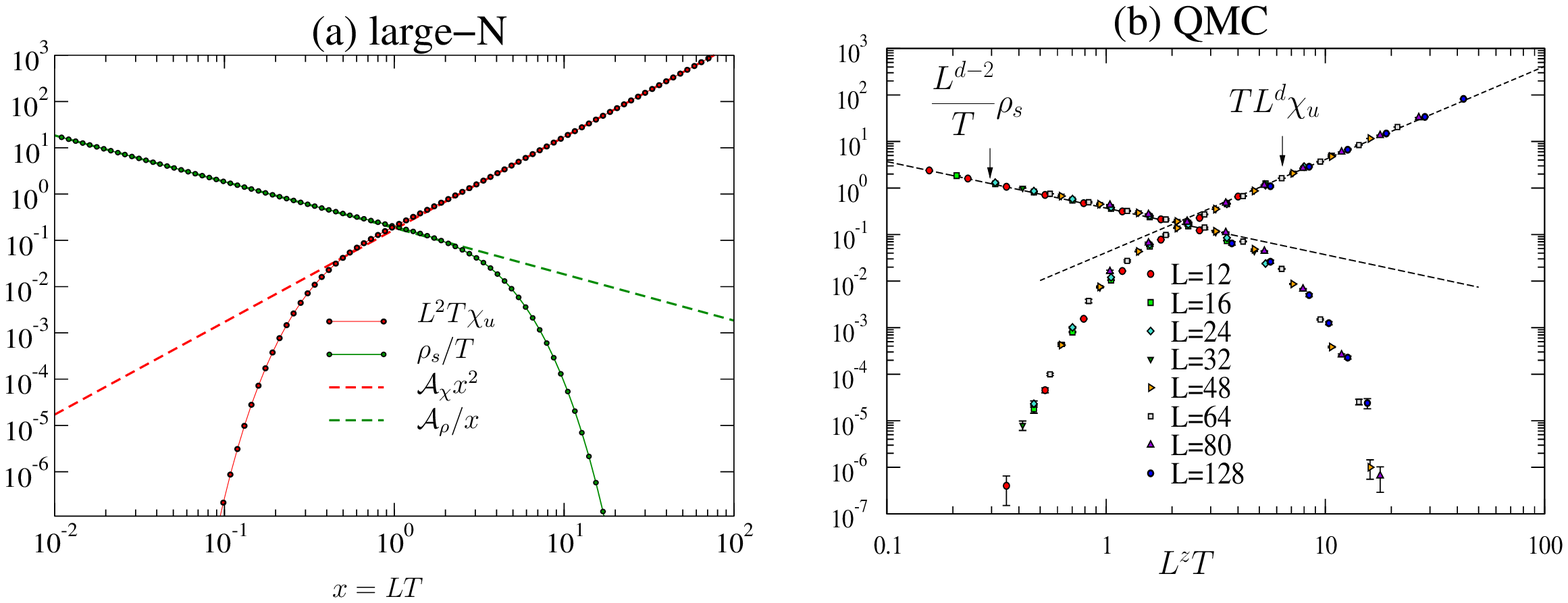}
\caption{ \label{fig:chirho} Universal finite-size scaling forms for $\chi_u$ and $\rho_s$. (a) for the $\mathbb{CP}^{N-1}$ field theory, evaluated in the $N=\infty$ limit, as described in the text. (b) from quantum Monte Carlo on the JQ model close to the N\'eel-VBS transition 
[reproduced from Ref.~\onlinecite{melkokaul}] . The functions plotted correspond to $\mathbb{Y}(x,0)$ and $\mathbb{Z}(x,0)$. There is no scale factor on the y-axis of this plot, but the x-axis has to be scaled properly with a non-universal velocity to make a comparison. }
\end{figure}

The computation of $\rho_s$ follows in exactly the same way, though with space and time interchanged. Again we can extract the amplitude, defined by the limiting behavior $\mathbb{Y}(x\rightarrow 0,0)=\mathcal{A}_\rho/x$, by converting sums into integrals, we find $\mathcal{A}^0_\rho/N=0.0926013$. The universal functions $\mathbb{Y}(x,0), \mathbb{Z}(x,0)$ can be evaluated numerically and are shown in Fig.~\ref{fig:chirho}.

{\em Comparison with JQ model: } The corresponding finite-size scaling functions for $\rho_s$ and $\chi_u$ for the JQ model have been computed before in Ref.~\onlinecite{melkokaul}. They are reproduced here in Fig.~\ref{fig:chirho}(b). There is a qualitative agreement between the $N=\infty$ calculations [Fig.~\ref{fig:chirho}(a)] and the numerical data. It is encouraging that the numerical data also shows the correct asymptotic forms for the scaling functions at large $LT$ for $\chi_u$ and small $LT$ for $\rho_s$. In order to avoid the complication of determining the non-universal velocity, $c$, it is useful to consider ratios of numbers where this quantity cancels. A completely universal number can be constructed by estimating $\mathcal{R}\equiv\mathcal{A}_\rho\sqrt{\mathcal{A}_\chi}$. At $N=\infty$ it takes the value: $\mathcal{R}_{N=\infty}=0.076642$. The quoted value of this combination of amplitudes in Ref.~\onlinecite{melkokaul} is 0.075(4). The normalizations of the quantities $\rho_s$ and $\chi_u$ in the large-N and QMC analysis is presented in Appendix~\ref{appendix:norm}. The agreement is surprisingly good. We can go a step ahead and compare the amplitudes directly. In order to do so, we need an estimate for the non-universal velocity. One way to estimate this quantity is to study the data in Fig.~\ref{fig:chirho}. Indeed, as is clear from the study in the Appendix, when $LT=c$ the system is perfectly cubic, and  hence the two universal functions plotted must be equal, i.e. $c$ is the value of $LT$ when the functions cross. By analyzing our data, we find: $c=2.4(3)$. Using this value of $c$, the QMC estimates for the amplitudes are $\mathcal{A}_\chi=0.23(6)$ and $\mathcal{A}_\rho=0.15(2)$ (using the data from Ref.~\onlinecite{melkokaul}), in reasonable agreement with the $N=\infty$ estimates ($\mathcal{A}^0_\chi=0.171$ and $\mathcal{A}^0_\rho=0.185$).

It would be clearly be  interesting to verify that the next $1/N$ correction to this quantity is actually small. In the next section we study a quantity for which we have succeeded in calculating these corrections. 

\section{Wilson Ratio}
\label{sec:wilson}
In the previous section we restricted ourselves to a $N=\infty$ calculation. To go beyond a qualitative discussion and actually compare numbers, it is essential to include $1/N$ corrections. We turn to this computation in the present section. It is also clearly of interest to focus on amplitude ratios which do not depend on the non-universal velocity, $c$. One such ratio is the so-called Wilson Ratio,
\begin{equation}
\mathcal{W}\equiv\frac{T\chi_u}{C_v}\approx \frac{\mathcal{A}_{\chi}}{\mathcal{A}_{C_V}}
\end{equation}
 where the second equality make use of the scaling forms, Eq.~(\ref{eq:scaling}) when $L\gg1/T$. Note that unlike $\mathcal{R}$ which requires an amplitude from the $LT\ll 1$ limit, $\mathcal{W}$ is a ratio of thermodynamic quantities, and is hence accessible even to possible experimental measurements.

At $N=\infty$ we can obtain the value the of both amplitudes analytically from the free energy, Eq.~(\ref{eq:freebox}), allowing an estimate of the Wilson ratio at $N=\infty$: 
\begin{equation}
\mathcal{W}_{N=\infty}=\frac{\mathcal{A}^0_\chi}{\mathcal{A}^0_{C_v}}\approx\frac{N( 0.0856271)}{N (1.83661)}
\approx 0.0466224
\end{equation}

Computations of the $1/N$ corrections are rather technical, requiring tedious analytic and numerical evaluations. The basic calculation has been set up in Sec.~\ref{sec:lgN}. The free energy has to be computed at next to leading order at finite-$T$, but in the thermodynamic limit. At order $1/N$, it receives contributions from Gaussian fluctuation of both the Lagrange multiplier $\lambda$ and the gauge field $A_\mu$. The resulting correction to the free energy, Eq.~(\ref{eq:free_en1}), has to be evaluated numerically and then numerical derivatives give the specific heat and susceptibility. Explicit details of these calculations may be found in Ref.~\onlinecite{kaulsachdev}, we will be content with only presenting the results here.
\begin{equation}
\lim_{N\rightarrow 2}\mathcal{W}_{\mathcal{O}(1/N)}=\frac{\mathcal{A}^{0}_{\chi}+\mathcal{A}^{1\lambda}_{\chi}+\mathcal{A}^{1A}_{\chi}}{\mathcal{A}^{0}_{C_V}+\mathcal{A}^{1\lambda}_{C_V}+
\mathcal{A}^{1A}_{C_V}}\approx \lim_{N\rightarrow 2} \frac{N(0.08562)-0.02650 + 0.26106}{N(1.8366)-0.38368+2.9928} \approx 0.0645
\end{equation}
where the superscript $0$ indicates $N=\infty$ values (these contributions are proportional to $N$ and evaluated at $N=2$), and the superscript $1\lambda$ and $1A$ indicate the leading $1/N$ correction from the $\lambda$ and $A_\mu$ fields (these contributions have no $N$ dependence). It should be noted that the gauge field fluctuations do produce rather large corrections to the $N=\infty$ amplitudes individually (unlike the $\lambda$ terms), so it is unclear how reliably they estimate the role of fluctuations. A proper estimate likely requires the inclusion of further terms in the expansion. It is re-assuring however that the $1/N$ corrections do have the correct sign for both $\mathcal{A}_\chi$ and $\mathcal{A}_\rho$, with respect to the QMC results (see Table~\ref{amptable}).

\begin{figure}[]
\includegraphics[width=5.in]{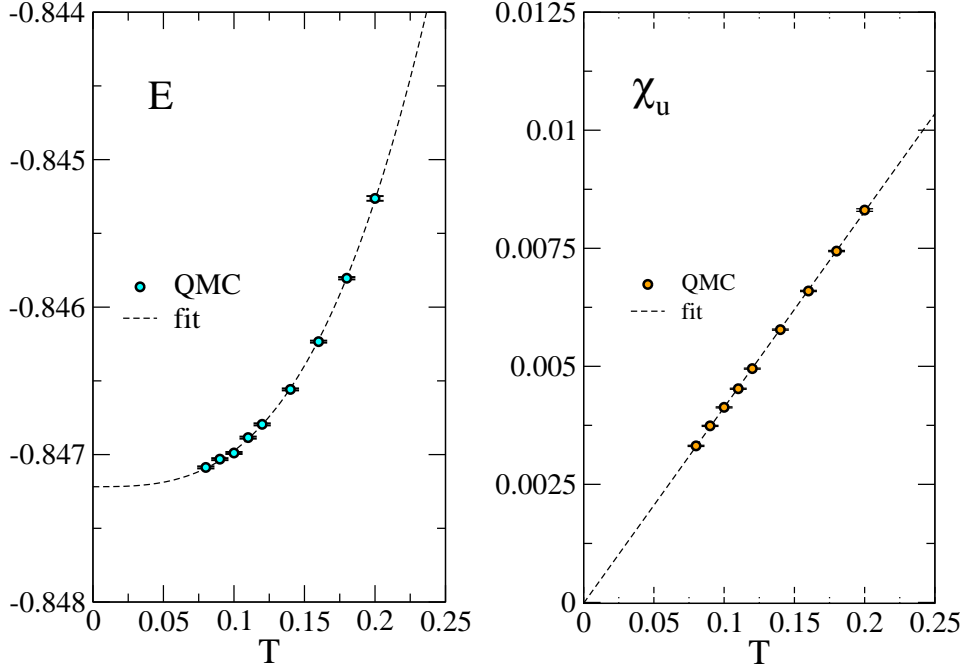}
\caption{\label{fig:enersus} Quantum critical scaling of the energy, $E$ and the uniform susceptibility, $\chi_u$ in the thermodynamic regime, $T\gg 1/L$. The size has been fixed to $L=128$ in these simulations, the data is converged to its infinite size limit within our error bars. Fits of $E(T)$ to the form $a_E+b_ET^3$ and $\chi_u(T)$ to the from $a_\chi +b_\chi T$ are shown as dashed lines. The chi-square merit function for these fits per degree of freedom (7 DOF in this case) are 1.91 and 1.96 respectively. From these fits we estimate, $\mathcal{A}_\chi/c^2=0.041(2)$ and $\mathcal{A}_{C_V}/c^2=0.75(5)$.}
\end{figure}

{\em Comparison with JQ model:} We extract the amplitudes for $C_V$ and $\chi_u$ by studying the finite temperature data on a $128\times 128$ system, close to the phase transition in the JQ model, we use $J/Q=0.038$. Because the specific heat requires a subtraction of two estimators, it turns out to be quite noisy, we hence find it preferable to look at the temperature dependence of the average energy, which can be measured very accurately. The QMC data and fits to it are shown and described in the inset of Fig.~\ref{fig:enersus}. From these fits, we can extract the Wilson Ratio. $\mathcal{W}_{\rm QMC}=0.055(5)$. This number is already fairly close to $\mathcal{W}_{N=\infty}=0.0466$, the mean-field value. 

\section{Conclusions}

In this paper, we have studied
the $\mathbb{CP}^{N-1}$ field theory at large-$N$ -- the fixed point at $N=\infty$ is known to be stable at large but finite $N$. The proof of existence (or non-existense) of a fixed point for the case of interest, $N=2$,  is highly non-trivial and does not yet exist. The large-$N$ limit, however, allows a controlled expansion of universal amplitudes and scaling functions that can be extrapolated to the case of interest, $N=2$. In particular, we have studied a number of amplitudes and scaling functions related to finite size and finite temperature effects; two ``universal'' amplitude ratios that can be extracted without the knowledge of $c$ were also studied, $\mathcal{R}=\mathcal{A}_\rho\sqrt{\mathcal{A}_\chi}$ and the Wilson Ratio $\mathcal{W}=\mathcal{A}_\chi/\mathcal{A}_{C_V}$. The Wilson ratio is of greater interest since it is well defined in the thermodynamic limit (allowing for instance a possible comparison with experiment). 

As argued in Ref.~\onlinecite{DQCP12}, the low energy behavior close to a transition between the N\'eel and VBS phases is expected to be described by the non-compact $\mathbb{CP}^1$ field theory. It is hence interesting to compare our large-N results with the analysis of the JQ model as a test for quantum criticality in the JQ model. To facilitate such a comparison, we have estimated a number of the equivalent universal numbers mentioned in the previous paragraph from quantum Monte Carlo simulations on the JQ model.
We have provided a catalogue of our estimates of these amplitudes in Table.~\ref{amptable}. The qualitative agreement for the finite-size scaling functions of $\chi_u$ and $\rho_s$ in Fig.~\ref{fig:chirho} and the quantitative comparison amplitude ratios studied here is encouraging; all the amplitude ratios at mean field agree reasonably with the QMC data. A fully convincing demonstration would require direct simulations of an appropriate discretization of the $\mathbb{CP}^1$ field theory on large lattices and comparison with the numerical values we have provided here. This is an exciting direction for future work.

\begin{table}[t]
\begin{spacing}{1.5}
\centering
\begin{tabular}{||c||c|c|c||} \hline\hline
 &Definition& QMC  & large-$N$ for $N=2$  \\
 \hline\hline
$\mathcal{A}_\chi$ & $\lim_{L\rightarrow \infty}\chi_u = (\mathcal{A}_\chi/c^2) T$ &0.23(6) & 0.17125  \\ 
\hline 
$\mathcal{A}_\rho$ & $\lim_{T\rightarrow 0}\rho_s = c\mathcal{A}_\rho/ L$ &0.15(2)  & 0.18520 \\ 
\hline 
$\mathcal{A}_{C_v}$ & $\lim_{L\rightarrow \infty}C_v = (\mathcal{A}_{C_V}/c^2) T^2$&4.3(3)& 3.6733 \\ 
\hline 
$\mathcal{W}$ & $\mathcal{A}_\chi/\mathcal{A}_{C_v}$&0.55(5) & 0.46622  \\ 
\hline 
$\mathcal{R}$ & $\mathcal{A}_\rho \sqrt{\mathcal{A}_\chi}$&0.075(4) & 0.076642  \\ 
\hline\hline
\end{tabular}
\end{spacing}
\caption{Table of amplitudes and amplitude ratios. The first column are the amplitudes (ratios) with error bars determined from Quantum Monte-Carlo as detailed in the text. Note that the amplitude ratios, $\mathcal{R,W}$ do not require an estimate of the non-universal velocity, $c$. The second column is obtained by setting $N=2$ in the $N=\infty$, $\mathbb{CP}^{N-1}$ saddle point theory; these values are known in principle with arbitrary accuracy.}
\label{amptable}
\end{table}

\section{Acknowledgements}

We are grateful to Subir Sachdev for collaboration on related work and both him and Michael Levin for a number of invaluable discussions. RKK acknowledges financial support from NSF DMR-0132874, DMR-0541988 and DMR-0537077.

\appendix

\section{Normalization of $\rho_s$ and $\chi_u$}
\label{appendix:norm}

In this appendix we provide a summary of how we have defined $\rho_s$ and $\chi_u$ both in the QMC calculations and the large-N expansion. The normalization has to be done properly to make a numerical comparison between the large-N and quantum Monte Carlo.

We define $\chi_u$ as the response to a uniform twist along the temporal direction.
\begin{equation}
\chi_u = \frac{\partial^2 f_\theta}{\partial \theta^2}
\end{equation}
where $\theta$ is the angle of the twist along the $z-$direction of the spin per unit of imaginary time and $f_\theta$ is the free energy per unit volume calculated with the imposed twist. It is easy to see that this definition reproduces the familiar meaning of $\chi_u$:
\begin{eqnarray}
V f_\theta & =& - T \ln Z_\theta = -T \ln {\rm Tr}[e^{-\epsilon H}e^{i\theta \Delta \tau S^z_{\rm tot}} \cdots] = -T \ln {\rm Tr} [ e^{-\beta H}e^{i \theta \beta S^z_{\rm tot}}]\\
\chi_u &=& \frac{\partial^2 f_\theta}{\partial \theta^2} = \frac{1}{V T}\langle \left (S^z_{\rm tot}\right )^2 \rangle
\end{eqnarray}
where we have used the fact that $[S^z_{\rm tot},H]=0$.

Now we can define $\rho_s$ in exactly the same way, as the response to a uniform twist along one spatial direction, say $x$,
\begin{equation}
\label{eq:rhodef}
\rho_s = \frac{\partial^2 f_\phi}{\partial \phi^2}
\end{equation}
where $\phi$ is the angle of the twist along the $z-$direction of the spin per unit length of space and $f_\phi$ is the free energy per unit volume calculated with the imposed twist. It is clear that the twisted partition function must be periodic in $\phi L$, {\em i.e.} $Z(\phi)= \sum_W Z_W e^{iW\phi L}$ (for exactly the same reason that the partition function of charged particles on a ring are periodic in the flux that threads the ring), where $W$ is summed on integers and is the winding number of the trajectories of the bosons that one would obtain by interpreting our spin model as hard-core bosons. Then by applying the formula Eq.~(\ref{eq:rhodef}), we arrive at the classic result,
\begin{equation}
\rho_s = T \langle W^2 \rangle,
\end{equation}
where $W$ is the so-called spatial winding number. For the full original derivation of this idea, see Ref.~\onlinecite{stiff1} and for an adaptation to the SSE method used here, see Ref.~\onlinecite{stiff2}.

Both these quantities can be calculated by imposing a similar twist to the $\mathbb{CP}^1$ field theory which results in modifying either the temporal(spatial) derivative as the case may be,
\begin{eqnarray}
\partial_\tau z_\alpha &\rightarrow& \partial_\tau z_\alpha  + i \theta \frac{\sigma^z_{\alpha\beta}}{2} z_\beta\\
\partial_x z_\alpha &\rightarrow& \partial_x z_\alpha  + i \phi \frac{\sigma^z_{\alpha\beta}}{2} z_\beta
\end{eqnarray}
The stiffness or susceptibility is then evaluated as the second derivative of the twisted free energy. Such a procedure has been carried out in the body of the text. 
 
\bibliographystyle{apsrev}


%
%
\end{document}